\documentstyle[12pt]{article}

\catcode`\@=11
\def\marginnote#1{}
\newcount\hour
\newcount\minute
\newtoks\amorpm
\hour=\time\divide\hour by60
\minute=\time{\multiply\hour by60 \global\advance\minute
by-\hour}\edef\standardtime{{\ifnum\hour<12
\global\amorpm={am}%
        \else\global\amorpm={pm}\advance\hour by-12 \fi
        \ifnum\hour=0 \hour=12 \fi
        \number\hour:\ifnum\minute<10
0\fi\number\minute\the\amorpm}}
\edef\militarytime{\number\hour:\ifnum\minute<10
0\fi\number\minute}

\def\draftlabel#1{{\@bsphack\if@filesw {\let\thepage\relax
   \xdef\@gtempa{\write\@auxout{\string
      \newlabel{#1}{{\@currentlabel}{\thepage}}}}}\@gtempa
   \if@nobreak \ifvmode\nobreak\fi\fi\fi\@esphack}
        \gdef\@eqnlabel{#1}}
\def\@eqnlabel{}
\def\@vacuum{}
\def\draftmarginnote#1{\marginpar{\raggedright\scriptsize\tt#1}}
\def\draft{\oddsidemargin -.5truein
        \def\@oddfoot{\sl preliminary draft \hfil
        \rm\thepage\hfil\sl\today\quad\militarytime}
        \let\@evenfoot\@oddfoot \overfullrule 3pt
        \let\label=\draftlabel
        \let\marginnote=\draftmarginnote
\def\@eqnnum{(\theequation)\rlap{\kern\marginparsep\tt\@eqnlabel}%
\global\let\@eqnlabel\@vacuum}  }

\def\numberbysection{\@addtoreset{equation}{section}
        \def\theequation{\thesection.\arabic{equation}}}
\def\underline#1{\relax\ifmmode\@@underline#1\else
 $\@@underline{\hbox{#1}}$\relax\fi}

\catcode`@=12
\relax

\numberbysection

\advance \topmargin by -\headheight
\advance \topmargin by -\headsep

\evensidemargin \oddsidemargin
\def\rf#1{(\ref{eq:#1})}
\def\lab#1{\label{eq:#1}}
\def\br{\begin{eqnarray}}
\def\er{\end{eqnarray}}
\def\be{\begin{equation}}
\def\ee{\end{equation}}

\def\({\left(}
\def\){\right)}

\relax

\begin{document}
\newcommand{\bi}[1]{\bibitem{#1}}
\begin{center}
\noindent {\Large{\bf{Nonminimal coupling, equivalence principle
and  exact Foldy-Wouthuysen transformation }}}
\end{center}
\vskip 3cm

\begin{center}
{{\bf {A. Accioly$^{a}$\footnote{Corresponding author.\\ {\it
E-mail address}: accioly@ift.unesp.br (A. Accioly)}, H.
Blas$^{a}$}, H. Mukai$^{b}$}}

 \vskip 0.3cm

 $^{a}${\it Instituto de F\'{\i}sica Te\'{o}rica,
Universidade Estadual Paulista,}\\ {\it Rua Pamplona 145,
01405-900  S\~{a}o Paulo, SP, Brazil}\\ $^{b}$ {\it Departamento
de F\'{\i}sica, Funda\c c\~ao Universidade Estadual de
Maring\'a,}\\ { \it Av. Colombo 5790, 87020-900 Maring\'a, PR,
Brazil}

\vskip 2cm
 \noindent{\large{\bf Abstract}}
\end{center}
\vskip 0.3cm It is shown that the  exact Foldy-Wouthuysen
transformation for spin-$0$ particles on spacetimes described by
the metrics $ds^2 = V^2 dt^2 - W^2 d {\bf{x}}^2$, where
$V=V({\bf{x}})$ and $W=W({\bf{x}})$, only exists if the scalar
field is nonminimally coupled to the Ricci scalar field with a
coupling constant equal to $\frac{1}{6}$. The nonminimal coupling
term, in turn, does not violate the equivalence principle. As an
application we obtain the nonrelativistic Foldy-Wouthuysen
Hamiltonian concerning the general solution to the linearized
field equations of higher-derivative gravity for a static
pointlike source in the Teyssandier gauge.

\vskip 1cm

\newpage

\section{Introduction}

\vskip 0.3cm

The COW experiment \cite{b1} as well as the Bonse-Wroblewski one \cite{b2} not
only shed a new light on the physical phenomena in which both
gravitational and quantum effects are interwoven but also showed
that the aforementioned phenomena are nomore beyond our reach. The
theoretical analysis concerning these experiments consisted simply
in inserting the Newtonian gravitational potential into the
Schr$\ddot{\rm{o}}$dinger equation. To improve this analysis we
need to learn, certainly, how to obtain an adequate interpretation
for relativistic wave equations in curved space. In other words,
we have to acquaint ourselves with the issue of the gravitational
effects on quantum mechanical systems. This can be done by
constructing the Foldy-Wouthuysen transformation (FWT)\cite{b3, b4} - the
keystone of relativistic quantum mechanics - for both bosons and
fermions coupled to the spacetime metric. Here we address
ourselves to the problem of finding the exact FWT for spin-0
particles coupled to the static metrics
\br
\lab{f1}
ds^2 = V^2 dt^2 - W^2 d{\bf{x}}^2\;\;\; ,
\er
where $V=V({\bf x})$ and $ W= W ({\bf x})$.

We employ  natural units in which $c = \hbar =1$.  Our convention
 are to use a metric with  signature  $(+ - - - )$ and define the
 Riemann and Ricci tensors as
  $R^\rho_{\;\; \lambda \mu \nu} = -
\partial_\nu \Gamma^\rho_{\;\; \lambda \mu}+ \partial_\mu
\Gamma^{\rho}_{\;\;\lambda \nu}$ $ - \Gamma^{\sigma}_{\;\;\lambda
\mu} \Gamma^{\rho}_{\;\;\sigma \nu} +
\Gamma^{\sigma}_{\;\;\lambda\nu} \Gamma^{\rho}_{\;\;\sigma\mu}$
and $R_{\mu\nu} = R^\rho_{\;\; \mu\nu \rho}$.

\vskip 1 cm

\section{Exact FWT for spin $0$ particle}

\vskip .4cm

The propagation of a free scalar $\phi$ with mass $m$ in Minkowski
spacetime is described by the Klein-Gordon (KG)
\begin{equation}
\lab{f2}
\left( \eta^{\mu\nu} \partial_{\mu} \partial_{\nu} + m^2 \right)
\phi = 0  \;\;\; ,
\end{equation}
where $\eta_{\mu\nu}$ is the Minkowski metric.

A generalization of \rf{f2} to a curved background is
\begin{equation}
\lab{f3}
\left( g^{\mu\nu} \nabla_{\mu} \nabla_{\nu} + m^2 + \lambda R
\right)  \phi = 0 \;\;\; ,
\end{equation}
where the numerical constant $\lambda$ represents the
direct coupling between the field and the curvature. Since the
coupling constant $\lambda$ can have any real value, we are on the
horns of a dilemma: Do we pick out the value $\lambda =0$, which
means that the field and the curvature are minimally coupled, or
do we choose $\lambda \neq 0$, implying in a nonminimal coupling
of the field to the curvature? Fortunately, there are two strong
arguments that seems to favour the choice $\lambda = \frac{1}{6}$:

{\it 1.} The equation for the massless scalar field is
conformally invariant\cite{b5}-\cite{b7}.

{\it 2.} Under the assumptions that {\it (i)} the scalar field satisfies \rf{f3}, and {\it (ii)} the field $\phi$ does
not violate the equivalence principle, $\lambda$ is forced to
assume the value $\frac{1}{6}$  \cite{b8}-\cite{b10}.

We assume thus that $\phi$ satisfies  the covariant KG equation
$$ \left( {\atop} \Box + m^2 + \frac{1}{6} R  {\atop}\right)\phi =
0 \;\; , $$
where
$$ \Box \equiv g^{\mu\nu} \nabla_{\mu}\nabla_{\nu} =
\frac{1}{\sqrt{-g}}\partial_{\mu} \left( \sqrt{-g}
g^{\mu\nu}\partial_{\nu} \right)\;\;\; . $$

Specializing now to the static metrics \rf{f1} we find
$$\ddot{\phi}  - F^2 \nabla^2 \phi - F^2{\boldmath{\nabla}}\ln
(VW) \cdot {\boldmath{\nabla}}\phi + m^2 V^2 \phi + \frac{1}{6} R
\phi^2 =0 \;\;\; , $$
where
$$ R = \frac{2}{W^4} \left( {\boldmath{\nabla}} W \right)^2 -
\frac{2}{V W^3} {\boldmath{\nabla}} V \cdot {\boldmath{\nabla}}W -
\frac{2}{V W^2} \nabla^2 V - \frac{4}{W^3}\nabla^2 W \;\;\; . $$

Here $F^2 \equiv \frac{V^2}{W^2}$ and the
differentiation with respect to time is denoted by dots.

In order to understand the physics of the equation above we write
it in first order form
$$ i \dot{\Phi} = {\cal{H}} \Phi \;\;\; , $$
where
\begin{eqnarray}
 \Phi &=& \left(
\begin{array}{c}
  \phi_1  \\
  \phi_2
\end{array}\right) \;\;\; , \nonumber\\
\phi_1 &=& \frac{1}{2} \left( {\atop} \phi + \frac{i}{m}
\dot{\phi} {\atop} \right) \;\; ,\,\, \phi_2 \,=\,
\frac{1}{2} \left( {\atop} \phi - \frac{i}{m} \dot{\phi}{\atop}
\right)\;\;\; ,\nonumber
\end{eqnarray}
with the Hamiltonian given by
\begin{equation}
\lab{f4}
{\cal{H}} = \frac{m}{2} \xi^T - \xi \theta \;\;\; ,
\end{equation}
where
\begin{eqnarray}
\xi&=& \left(
\begin{array}{cc}
  1 & 1 \\
 -1 & -1
\end{array}\right)\;\;\; , \nonumber\\
 \theta &= & \frac{F^2}{2m} \nabla^2 - \frac{F^2}{2m}
{\boldmath{\nabla}} \ln (VW) \cdot {\boldmath{\nabla}} -
\frac{m}{2} V^2 - \frac{1}{24 m}V^2 R \;\;\; . \nonumber
\end{eqnarray}

The matrix $\xi$ has the following algebraic properties
$$ \xi^2 = 0 \;\;\; , \;\;\; \left\{ \xi , \xi^{T} \right\} = 4
\;\;\; .$$

Redefining the scalar field and the Hamiltonian
 $$ \Phi' = f \Phi \;\;\; ,\;\;\;
{\cal{H}'}= f {\cal{H}} f^{-1} \;\;\; , $$
with
$$ f  = V^{-1/2} W^{3/2} \;\;\; , $$
we obtain a new Hamiltonian which is explicitly
Hermitian with respect to the usual flat space measure:
$$ {\cal{H}}' = \frac{m}{2} \xi^{T} - \xi
\theta ' \;\;\; , $$
where
\begin{eqnarray}
 \theta ' &=& f \theta f^{-1} \nonumber\\
 &=& - \frac{m}{2} V^2 - \frac{F}{2m} {\hat{{\bf p}}}^2 F +
 \frac{1}{8m} {\bf \nabla}F \cdot {\bf \nabla} F - \frac{1}{12m} F
 {\bf \nabla}^2 F \;\;\; .\nonumber
 \end{eqnarray}

Here $ \hat{\bf{p}} = - i {\boldmath{\nabla}}$ denotes the
momentum operator.

It is astonishing and at the same time fascinating that the square
of the transformed Hamiltonian ${\cal{H}}'$,
$$ {\cal{H}}^{\tiny{\Box}} = - \frac{m}{2} \theta ' \left\{ \xi
,\xi ^{T} \right\} = -2 m \theta ' I\;\;\; , $$
where
$$ I = \left(
\begin{array}{cc}
 1 & 0 \\
 0 &1
\end{array}\right) \;\;\; ,
$$
is diagonal and, obviously, an even operator. Formally,
we have
$$ \sqrt{{\cal{H}}^{\tiny{\Box}}} = (- 2m \theta ' )^{1/2} \eta
\;\;\; , $$
where
$$ \eta= \left(
\begin{array}{cc}
1 &0 \\
  0 & -1
\end{array} \right) \;\;\; , \;\;\; \eta^2 =1 \;\;\; .
$$

Accordingly, ${\cal H}' = f {\cal H} f^{-1} $ is the exact FWT for
the KG equation in curved space \cite{b11}.

Now, taking into account, that the Hamiltonian squared can be
rewritten as
$$ {\cal{H}}^{\tiny{\Box}} = m^2 V^2 + F \hat{p}^2 F -\frac{1}{4}
{\boldmath{\nabla}} F \cdot {\boldmath{\nabla}} F + \frac{1}{6} F
{\nabla}^2 F \;\;\; , $$
we promptly obtain the quasirelativistic Hamiltonian
simply  by assuming that the dominating term is $m$:
\begin{equation}
\lab{f5}
 \sqrt{{\cal{H}}^{\tiny{\Box}}} \approx   m V + \frac{1}{4m}
\left( W^{-1} {\hat{p}}^2 F + F {\hat{p}}^2 W^{-1} \right) -
\frac{1}{8 m V} {\bf{\nabla}} F \cdot {\bf{\nabla}}F + \frac{1}{12
m W } {\nabla}^2 F \;\;\; .
\end{equation}

Two  comments are in order here:

\noindent {\it (i)} The preceding  Hamiltonian contains the
gravitational Darwin term
$$ \frac{1}{12 m W} \nabla^2 F \;\;\; , $$
which is lacking in all the works on this subject \cite{b12}
with the exception of that by Obukhov [13]. \vskip 0.2 cm

\noindent{\it (ii)} \rf{f5} is identical to the spinless sector found
 in Ref.  \cite{b13} for the Dirac particle except for the Darwin term
which is one third of the corresponding term in the fermionic
case. Fermions and bosons, of course, lead to different Darwin
terms \cite{b14}.

\vskip 1 cm

\section{The nonrelativistic FW Hamiltonian concerning the general solution to the linearized field equations
of higher-derivative gravity for a static pointilike source in the
Teyssandier gauge }

\vskip .4cm

Higher-derivative gravity is defined by the action
$$ S= \int d^4 x \sqrt{-g} \left[ \frac{2 R}{\kappa^2} +
\frac{\alpha}{2} R^2 + \frac{\beta}{2} R_{\mu\nu}^2 \right] - \int
d^4 x \sqrt{-g} {\cal{L}}_{M} \;\;\; , $$

\noindent where ${\cal{L}}_M $ is the Lagrangian density for
matter, $\kappa^2 \equiv 32 \pi G$, with $G$ being the Newton's
constant, and $\alpha$ and $\beta$ are dimensionless parameters.
The corresponding field equations are
\begin{eqnarray}
\frac{2}{\kappa^2} G_{\mu\nu} &+& \frac{\alpha}{2} \left[
-\frac{1}{2} g_{\mu\nu} R^2 + 2 R R_{\mu\nu} + 2 \nabla_{\mu}
\nabla_{\nu} R - 2 g_{\mu\nu} \Box R {\atop} \right]\nonumber\\
&+& \frac{\beta}{2} \left[ -\frac{1}{2} g_{\mu\nu}
R^2_{\rho\sigma} + \nabla_{\mu} \nabla_{\nu} R + 2
R_{\mu\rho\lambda\nu} R^{\rho\lambda} - \frac{1}{2} g_{\mu\nu}
\Box R - \Box R_{\mu\nu} {\atop} \right] + \frac{1}{2} T_{\mu\nu}
= 0 \;\;\; ,
\lab{f6}
\end{eqnarray}

\noindent where $$ \delta \int d^4 x \sqrt{-g} {\cal{L}}_M \equiv
\int \sqrt{-g} d^4x \frac{T^{\mu\nu}}{2} \delta g_{\mu\nu} \;\;\;.
$$

In the weak field approximation, i.e., $$ g_{\mu\nu} =
\eta_{\mu\nu} + \kappa h_{\mu\nu} \;\;\; , $$

\noindent and in the Teyssandier gauge [15], \rf{f6} reduces to $$
\left( {\atop} 1 - \frac{\beta \kappa^2}{4} \Box \right) \left( -
\frac{1}{2} \Box h_{\mu\nu} + \frac{1}{6} \eta_{\mu\nu} \bar{R}
{\atop} \right) = \frac{\kappa}{4} \left[{\atop} T_{\mu\nu} -
\frac{1}{3} T \eta_{\mu\nu} {\atop} \right]\;\;\; ,$$

\noindent where

$$ \bar{R}= \frac{1}{2} \Box h - \gamma^{\mu\nu}_{\;\;\;,\mu\nu}
\;\;\;,$$

\noindent with $\gamma_{\mu\nu} = h_{\mu\nu} - \frac{1}{2}
\eta_{\mu\nu} h$. Here indices are raised (lowered) using $
\eta^{\mu\nu} (\eta_{\mu\nu})$. Teyssandier showed that the
general solution of the above equations is given by \cite{b15}

$$ h_{\mu\nu} = h_{\mu\nu}^{\;\;(E)} + \psi_{\mu\nu} - \Phi
\eta_{\mu\nu}\;\;\;, $$

\noindent where $h_{\mu\nu}^{(E)}, \;\psi_{\mu\nu}$and $\Phi$
satisfy the following equations:

\begin{eqnarray}
\Box h_{\mu\nu}^{(E)} &=& \frac{\kappa}{2} \left[ \frac{T
\eta_{\mu\nu}}{2} - T_{\mu\nu} {\atop} \right]
\;\;,\;\;\gamma_{\mu\nu}^{(E),\nu}=0 \;\;,\;\;
\gamma_{\mu\nu}^{(E)} =h_{\mu\nu}^{(E)} - \frac{1}{2}
\eta_{\mu\nu} h^{(E)}; \nonumber\\ \left( \Box + m_1^2 \right)
\psi_{\mu\nu} &=& \frac{\kappa}{2} \left[ {\atop} T_{\mu\nu} -
\frac{1}{3} T \eta_{\mu\nu} {\atop} \right] \;\;, \;\;
\psi^{\mu\nu}_{\;\;, \mu\nu} - \Box \psi =0\;\; ;\;\; \left( \Box
+ m_0^2 \right) \Phi = \frac{\kappa T}{12} \;\;\;. \nonumber
\end{eqnarray}

Here $m_0^2 \equiv \frac{2}{\kappa^2 \left( 3 \alpha +
\beta \right)} \;\;,\;\; m_1^2 \equiv - \frac{4}{\kappa^2 \beta}
\;\;\;.$

For a point particle of mass $M$ located at ${\bf r} = {\bf 0}$,
the general solution of the equations above is \cite{b16, b15, b17}

\begin{eqnarray}
g_{00} &=& 1 + 2 MG \left[ - \frac{1}{r} - \frac{1}{3}
\frac{e^{-m_0 r}}{r} + \frac{4}{3} \frac{e^{-m_1
r}}{r}\right]\;\;\;,\lab{f7}\\ 
\lab{f8}
g_{11} &=& g_{22} = g_{33}= - 1 + 2 MG
\left[ -\frac{1}{r} + \frac{2}{3} \frac{e^{-m_1 r}}{r} +
\frac{1}{3} \frac{e^{-m_0 r}}{r} \right] \;\;\;.
\end{eqnarray}

It is worth mentioning that we have assumed that $m_0^2$
and $m_1^2$ are real, in order to assure asymptotic agreement of
the theory with Newton's law.

From \rf{f7} and \rf{f8} we get immediately

\begin{eqnarray}
V &=& 1 + MG \left[ - \frac{1}{r} - \frac{1}{3} \frac{e^{-m_0
r}}{r} + \frac{4}{3} \frac{e^{-m_1 r}}{r}\right]\;\;\;,\lab{f9}
\\ 
\lab{f10}
W &=& 1-MG \left[ -\frac{1}{r} + \frac{2}{3} \frac{e^{-m_1 r}}{r} +
\frac{1}{3} \frac{e^{-m_0 r}}{r} \right] \;\;\;.
\end{eqnarray}

Inserting \rf{f9} and \rf{f10} into \rf{f5} we come to the conclusion that the
nonrelativistic FW Hamiltonian is given by
\br
\sqrt{{\cal{H}}^{\Box}} &=& \left\{ {\atop} m + \left( {\atop} 1 +
\frac{e^{-m_0 r}}{3} - \frac{4}{3} e^{-m_1 r} {\atop} \right) m \;
{\bf g} \cdot {\bf x} + \frac{\hat{{\bf{p}}}^2}{2m} + \frac{1}{2m}\hat{\bf{p}}.u\hat{\bf{p}} \right.
\nonumber\\
\lab{f11}
 &+& \left. \frac{1}{2m} {\bf \nabla} \cdot \left[ {\atop} \(\frac{7}{6} - \left( 1 + m_1 r
\right) e^{-m_1 r} - \frac{1}{6}\left( 1 + m_0 r
\right) e^{-m_0 r}\)  {\bf g} \right] \right\} \eta \;\;\; ,
\er
with
$$ {\bf g} = - GM \frac{{\bf r}}{r^3}\;\;\;$$ and $$u \equiv MG(-\frac{3}{r}+\frac{1}{3} \frac{e^{-m_{0} r}}{r}+ \frac{8}{3} \frac{e^{-m_{1} r}}{r})$$.

Note that \rf{f11} tends to
\begin{equation}
\lab{f12}
\sqrt{{\cal{H}}^{\Box}} =\left[ {\atop} m+ m \; {\bf g} \cdot
{\bf x} + \frac{\hat{{\bf p}}^2}{2m} + \frac{3}{2m} \hat{{\bf{p}}} \cdot({\bf g} \cdot {\bf x}) \hat{{\bf{p}}} + \frac{7}{12m} {\bf \nabla} \cdot {\bf g} \right] \eta \;\;\;,
\end{equation}
as $m_0 , m_1 \rightarrow  \infty$, which is nothing but
the nonrelativistic Hamiltonian for the spin-0 particle in the external gravitational field of a central gravitating body of mass $M$ in the
framework of ordinary general relativity. The first terms of this expansion has been found in \cite{Wajima} by means of another method. The coefficient in the last term of \rf{f12} differs from that of \cite{b13} since the Darwin term contribution in the scalar case enters as $1/3$ of the relevant term for fermions. It is worth mentioning
that we have neglected in \rf{f11} and \rf{f12} the higher order
relativistic and gravitational/inertial terms.

\vskip 1 cm

\section{The nonminimal coupling term does not
violate the equivalence principle }

\vskip .4cm

Let us now  show that the nonminimal coupling term does not
violate the equivalence principle for spin-0 particles by making a
comparison of the true gravitational coupling with the pure
inertial case. To do that, we recall that in the case of the flat
Minkowski space in accelerated frame,

$$ V= 1+ {\bf a}\cdot {\bf x} \;\;\; , \;\;\; W=1\;\;\;.$$

Consequently, the corresponding nonrelativistic FW Hamiltonian is
\begin{equation}
\lab{f13}
\sqrt{  {\cal{H}}^{\Box} } = \left[ {\atop} m + m \; {\bf a} \cdot
{\bf x} + \frac{\hat{{\bf{p}}}^2}{2m}+ \frac{1}{2m} \hat{{\bf{p}}}\cdot ({\bf a}\cdot {\bf x})\hat{{\bf{p}}} \right] \eta \;\;\; ,
\end{equation}
where we have also neglected the higher order
relativistic and gravitational/inertial terms.

For the particle $m$ far away from the body $M$ one can neglect the terms $\frac{3}{2m} \hat{{\bf{p}}} \cdot({\bf g} \cdot {\bf x}) \hat{{\bf{p}}}$\, and $\frac{1}{2m} \hat{{\bf{p}}}\cdot ({\bf a}\cdot {\bf x})\hat{{\bf{p}}}$ in \rf{f12} and \rf{f13}, respectively, since they are of the order $(GM)/(m r)$. In \rf{f13} we are assuming that ${\bf a}$ is such that $|{\bf a}\cdot {\bf x}|/m$ is of the order $(GM)/(m r)$. Then, we come to the conclusion that the covariant KG
 theory with $\lambda = \frac{1}{6}$ is in agreement with the
 equivalence principle.

 \vskip 1 cm

\section{Final remarks }

\vskip .4cm

We have shown that {\it (i)} the KG equation only admits exact FWT
if $\lambda = \frac{1}{6}$, and {\it(ii)} the nonminimal coupling
term, contrary to a previous claim \cite{lightman}, does not violate the equivalence
principle. We are not claiming, however, that $ \lambda =
\frac{1}{6}$ is the correct coupling for the various scalar
particles. The question of which value(s) of $\lambda$ should
constitute the correct coupling(s) to gravity depends on the particular quantum field theory used for the scalar field $\phi$ (see e.g. \cite{faraoni} and references therein).

 \vskip 1cm

\noindent {\large{\bf{Acknowledgement}}}

\vskip 0.2cm
 H. Blas  gratefully acknowledges  financial support
 from  Funda\c c\~ao de Amparo \`a
Pesquisa do Estado de S\~ao Paulo (FAPESP).

\end{document}